\documentclass[aps,twocolumn,groupedaddress,nofootinbib,prb,showpacs]{revtex4-1}
\usepackage{siunitx}
\usepackage{bm}
\usepackage{pgfplots}
\usepackage[utf8]{inputenc}
\usepackage{amsmath}
\usepackage{multirow}
\usepackage{comment}
\pgfplotsset{compat=1.14}
\newcommand{\cuo}{Cu$_{\text{2}}$O} % print Cu_2O in textmode
\newcommand{\sy}[2][+]{\Gamma^{#1}_{#2}} % print representation names (default with positive parity)
\newcommand{\oh}{\mathcal{H}}
\newcommand{\bk}{\bm{k}}
\newcommand{\br}{\bm{r}}
\newcommand{\bn}{\bm{n}}
\newcommand{\bi}{\bm{I}}
\newcommand{\bs}{\bm{\sigma}}
\newcommand{\be}{\bm{\epsilon}}
\newcommand{\ids}{\bi\cdot\bs}
\DeclareMathOperator{\Tr}{Tr}
\DeclareMathOperator{\Id}{Id}

\allowdisplaybreaks
\begin{document}

\title{Waveguides for Rydberg excitons in {\cuo} from strain traps}

\author{Sjard Ole \surname{Kr{\"u}ger}}
\email[]{sjard.krueger@uni-rostock.de}
\author{Stefan \surname{Scheel}}
\affiliation{Institut f{\"u}r Physik, Universit{\"a}t Rostock, 
Albert-Einstein-Stra{\ss}e 23, D-18059 Rostock, Germany}

\date{\today}

\begin{abstract}
We investigate the formation of waveguides for Rydberg excitons in {\cuo} from 
cylindrical stressors as alternatives to optical traps. We show that the 
achievable potential depths can easily reach the $\si{\milli\eV}$ and the trap 
frequencies the $\si{\giga\hertz}$ regimes. For Rydberg excitons, we find that 
it is sufficient to consider only the shift of the band gap, whereas the 
excitonic binding energies remain almost unchanged.
\end{abstract}

\maketitle

\section{Introduction}
\label{sec:intro}

Excitons, bound states of an electron and a positively charged hole, are the 
fundamental optical excitations in a semiconductor. Cuprous oxide ({\cuo}) was 
the first material in which the formation of excitons with principal quantum 
numbers up to $n=8$ was experimentally  demonstrated \cite{gross1952,gross1956}.
In recent years, the variety of observed excitonic states has increased 
greatly \cite{kazimierczuk2014,thewes2015,schoene2016_prb}, including highly 
excited states of principal quantum numbers up to $n=25$ and
orbital quantum numbers of $\ell = 5$. These highly excited states do already 
show all the hallmarks of the Rydberg blockade \cite{kazimierczuk2014} as well 
as signs of quantum coherence \cite{grunwald2016}.

Strain potentials from inhomogeneous strain fields alter the band structure of 
the crystal and can generate effective trapping potentials. They have been used 
and extensively investigated in the pursuit of exciton condensation in {\cuo} 
\cite{trauernicht1986,snoke2000,naka2002,naka2004,yoshioka2011,stolz2012,
schwartz2012}.
Due to the long lifetime of the 1S paraexciton of its yellow exciton series 
($\sim\si{\micro\second}$), most research has hitherto been focused on 
the yellow 1S para- and orthoexcitons. Strain influence on the yellow P-excitons 
has been investigated in thin films\cite{iwamitsu2014}. In comparison to the dipole traps 
frequently used in atomic physics, strain traps potentially offer a greater 
variety of achievable trap geometries, as the shape of the trap depends 
strongly on the shape of the stressor, the stress applied to it, the excitonic 
state in question as well as the orientation of the crystal relative to the 
stress \cite{naka2004,sandfort2011}. Furthermore, the achievable potential 
depths ($\sim\si{\milli\eV}$) and trap frequencies ($\sim\si{\giga\hertz}$) are 
much higher than those of the atomic dipole traps ($\sim \si{\micro\eV}$ and 
$\sim \si{\mega\hertz}$) \cite{grimm2000}.

Excitonic traps may be a useful experimental tool for the investigation
of exciton-exciton interactions and many-body effects in exciton populations. The 
1D potential landscapes induced by cylindrical stressors offer a positional control
over the exciton populations in two dimensions and could pave the way for the investigation
of 1D many-body interactions between excitons.

\begin{figure}
\includegraphics[width=\columnwidth]{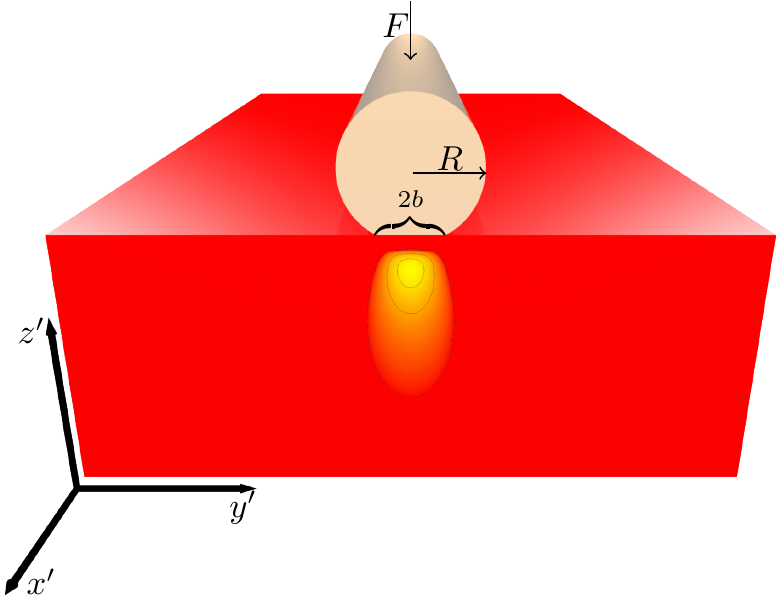}
\caption{Schematic cross section of the proposed geometry with a cylindrical 
stressor acting on a {\cuo} crystal with line load $F$ (force per unit length) 
and the resulting trapping potential. 
$b$ denotes the half width of the contact area.\label{fig:skizze}}
\end{figure}

In this article, we investigate the modification of the Rydberg 
exciton resonances in {\cuo} under the influence of strain from a cylindrical 
stressor (see Fig. \ref{fig:skizze}). The strain field $\be(\br)$ will be 
calculated using Hertzian contact theory \cite{mcewen1949,manghnani1974}, while 
the band Hamiltonian derived by Suzuki and Hensel \cite{suzuki1974} will be 
employed in order to describe the strain dependence of the hole motion, from 
which the band-gap shifts $\Delta E_g(\br)$ can be obtained. The shifts of the 
binding energies will be calculated using a two-band model similar to 
Ref.~\cite{schoene2016_prb}, generalised to anisotropic band structures.

The article is organised as follows. Section~\ref{sec:e-h-motion} introduces 
the band and strain Hamiltonians used in this study. In 
Sec.~\ref{sec:bandgap-shift} the strain fields and the 
resulting band-gap shifts will be calculated for certain geometries
and stresses. The influence on the excitonic binding energies will be discussed 
in Sec.~\ref{sec:binding-energies}, and Sec. \ref{sec:outlook} presents the 
conclusion and an outlook.

\section{Electron and hole motion under the influence of strain}
\label{sec:e-h-motion}

We begin with a brief review of the band structure of {\cuo} under the 
influence of strain. Without spin, the uppermost valence band of {\cuo} has 
$\sy{5}$-symmetry at the zone center, which splits into a twofold degenerate 
$\sy{7}$- and a fourfold degenerate $\sy{8}$-band under the influence of 
spin-orbit interaction. An effective $6\times 6$-Hamiltonian for these valence 
bands up to second order in the hole momentum $\hbar\bk_h$ and first order in 
the components of the strain tensor $\epsilon_{ij}$ ($i,j = x, y, z$) can be 
constructed based on group-theoretical considerations \cite{suzuki1974}.
It takes the form $\oh_v(\bk_h, \be) = \oh_{so} + \oh_{\bk_h} + \oh_{\be}$ 
where $\oh_{so}=-\frac{\Delta}{3} (2 + \ids)$ describes the $\bk_h$-independent 
spin-orbit splitting, $\bi$ denotes the vector of the angular-momentum matrices 
for $I=1$, $\bs$ the vector of the Pauli matrices and $\Delta$ the spin-orbit 
splitting.
Note that our definition of the spin-orbit splitting
guarantees that the band edge of the $\sy{7}$ valence band is located at the 
origin of the energy scale. 

The $\bk_h$-dependent part of the Hamiltonian is given by
\begin{gather}
\oh_{\bk_h} =\frac{\hbar^2}{2m_e}\bigg\{ \left(A_1 + B_1 \ids\right)\bk_h^2 
\nonumber\\
+ \left[A_2\left(I_x^2 - \frac{1}{3}\bi^2\right) + B_2\left(I_x\sigma_x - 
\frac{1}{3}\ids\right)\right]k_{x,h}^2 + \mathrm{c.p.} \nonumber\\
+\left[A_3(I_xI_y + I_yI_x) + B_3(I_x\sigma_y+I_y\sigma_x)\right] 
\{k_{x,h}k_{y,h}\} + \mathrm{c.p.}\bigg\}
\nonumber\\ \label{eq:sh-k}
\end{gather}
where $\mathrm{c.p.}$ denotes cyclic permutations, 
$\{k_ik_j\}=(k_ik_j + k_jk_i)/2$ is the symmetric product of momenta, and the 
$A_i$ and $B_i$ are dimensionless material constants. 
Values for all material properties of {\cuo} used in this work can be found in Table \ref{tab:mat-prop}. 
In the absence of external electromagnetic fields, the symmetric, cartesian strain tensor $\be$ 
with components $\epsilon_{ij}$ transforms according to the same reducible 
representation of the point group $O_h$ as $k_ik_j$, hence the 
strain Hamiltonian for the valence band has an analogous form to $\oh_{\bk_h}$
\begin{gather}
\oh_{\be} =\left(D_1 + E_1 \ids\right) \Tr(\be) \nonumber\\
+ \left[D_2\left(I_x^2 - \frac{1}{3}\bi^2\right) + E_2\left(I_x\sigma_x - 
\frac{1}{3}\ids\right)\right]\epsilon_{xx} + \mathrm{c.p.} \nonumber \\
+\left[D_3\left(I_xI_y + I_yI_x\right) + E_3\left(I_x\sigma_y + 
I_y\sigma_x\right)\right]\epsilon_{xy} + \mathrm{c.p.} \label{eq:sh-eps}
\end{gather}
with the deformation potentials $D_i$ and $E_i$. The $E_i$ are assumed to be 
negligible, as is usually done in the literature \cite{naka2004,sandfort2011}.

The equivalent effective Hamiltonian for the $\sy{6}$ conduction band is given 
by the $2\times 2$-matrix
\begin{equation}
\oh_c(\bk_e, \be) = E_g + \frac{\hbar^2}{2m_e} A_4 \bk_e^2 + C_1 
\Tr(\be).\label{eq:ham-con}
\end{equation}
A spatially homogeneous strain preserves the inversion symmetry as well as the 
time-reversal symmetry. Thus, the twofold degeneracy over the whole Brillouin 
zone of both the $\sy{7}$ valence band and the $\sy{6}$ conduction band is 
maintained.
As long as the strain field is approximately constant across the exciton 
volume, the assumption of spatial homogeneity is justified.

The kinetic energy of the relative electron-hole motion $\oh_r(\bk,\be) = 
\oh_c(\bk,\be)\otimes\Id_v-\Id_c\otimes\oh_v(\bk, \be)$ for vanishing 
center-of-mass momentum $\hbar\bm{K}=0$ is thus given by the $12\times 12$ 
matrix
\begin{equation}
\oh_r(\bk,\be)=\Id_c \otimes(E_g-\oh_{v}(\bk,\be) )\label{eq:hr}
\end{equation}
with modified parameters $A_1 \mapsto A_1 - A_4$ and $D_1\mapsto D_1 - C_1 := 
D'_1$. Here, $\Id_c$ and $\Id_v$ denote the identity matrices in the basis of 
conduction-band and valence-band states, respectively.
This formulation is possible as $\oh_c(\bk_e, \be)$ is proportional to 
$\Id_c$ and the terms in $\oh_v(\bk_h, \be)$ containing $A_1$ and $D_1$ are 
proportional to $\Id_v$.

The excitonic strain potential can be broken down into a pure band-gap shift 
$\Delta E_g$ and a binding energy shift $\Delta E_b$ induced by the deformation 
of the bands (for nS orthoexcitons not considered here, there is an additional 
strain-dependent contribution from the exchange interaction \cite{naka2004}). 
The band-gap shift $\Delta E_g$ can be evaluated in third-order perturbation 
theory as
\begin{widetext}
\begin{gather}
\Delta E_g (\be) = -D'_1 \Tr(\be) %\nonumber\\
- \frac{2D_3^2 \left[\epsilon_{xy}^2+\epsilon_{yz}^2 +\epsilon_{zx}^2 
\right]}{3 \Delta}- \frac{D_2^2 \left[(\epsilon_{xx} - \epsilon_{yy})^2 
+(\epsilon_{yy} - \epsilon_{zz})^2+(\epsilon_{zz} - \epsilon_{xx})^2 \right] }{9 
\Delta}
\nonumber\\
-\frac{D_2^3(\epsilon_{xx} + \epsilon_{yy} - 2\epsilon_{zz})(\epsilon_{xx} -2 
\epsilon_{yy} +\epsilon_{zz})(-2\epsilon_{xx} + \epsilon_{yy} 
+\epsilon_{zz})}{27\Delta^2} + \frac{2D_3^3 
\epsilon_{xy}\epsilon_{yz}\epsilon_{zx}}{\Delta^2}
\nonumber\\
+ \frac{D_2D_3^2~\left[\epsilon_{xx} (\epsilon_{xy}^2 + \epsilon_{zx}^2 - 2 
\epsilon_{yz}^2) + \epsilon_{yy} (\epsilon_{xy}^2 + \epsilon_{yz}^2 - 2 
\epsilon_{zx}^2) + \epsilon_{zz} (\epsilon_{zx}^2 + \epsilon_{yz}^2 - 2 
\epsilon_{xy}^2)\right]}{3\Delta^2}.
\label{eq:perturbation}
\end{gather}
\end{widetext}
For a compressive strain with $\Tr(\be) < 0$, the first-order term raises 
the band gap and thus results in a repulsive contribution to the excitonic 
potential while the second-order terms lower the band gap.

\begin{table}
\def\arraystretch{1.5}
\caption{Material properties of {\cuo} used in this work.\label{tab:mat-prop}}
\begin{ruledtabular}
\begin{tabular}{r  l l | r l l}
$D'_1$ & $2.1~\si{\eV}$& \multirow{3}{*}{\cite{trebin1981}}&$A_1$&$-1.76$&
\multirow{6}{*}{\cite{schoene2016_prb}}\\
$D_2$ & $-3.9~\si{\eV}$ &&$A_2$&$4.519$&\\
$D_3$ & $1.9~\si{\eV}$&&$A_3$&$-2.201$& \\\cline{1-3}
$S_{11}$ & $41.69~\si{\per\tera\Pa}$ &\multirow{3}{*}{\cite{manghnani1974}}
&$B_1$&$0.02$&\\
$S_{12}$ & $-19.36~\si{\per\tera\Pa}$&&$B_2$&$-0.022$& \\
$S_{44}$ & $82.64~\si{\per\tera\Pa}$ &&$B_3$&$-0.202$& \\\hline
$\varepsilon_r$ & $7.5$ & \cite{carabatos1968}& $\Delta$&$131~\si{\milli\eV}$
&\cite{gross1956}\\\hline
$A_4$ & $1.01$ & \cite{hodby1976}&$E_g$&$2.17208~\si{\eV}$&\cite{kazimierczuk2014}\\
\end{tabular}
\end{ruledtabular}
\end{table}

\section{Band-gap shifts under the influence of strain from a cylindrical 
stressor}
\label{sec:bandgap-shift}

For some geometries, the stress field can be calculated analytically using 
Hertzian contact theory. It should be noted here that Hertzian contact theory 
assumes elastically isotropic materials. However, comparisons between 
finite-element calculations and Hertzian results for germanium (whose anisotropy 
is stronger than that of {\cuo}) have shown satisfying agreement 
\cite{markiewicz1977}.

In the case of an infinitely long cylindrical stressor 
in contact with a planar surface along the $x'$ axis of the $(x',y')$-plane 
(see Fig.~\ref{fig:skizze}), the components of the stress tensor 
$\sigma_{i'j'}$ are 
given by \cite{mcewen1949}$^{,}$\footnote{The original paper apparently 
contains a sign error in the definition of $\sigma_{x'x'}$ as the equations 
given there do not describe a plane strain, as claimed ($\epsilon_{x'x'} = 
Y^{-1}\left[\sigma_{x'x'} -\nu(\sigma_{y'y'} + \sigma_{z'z'})\right]$ for 
isotropic materials).}
\begin{gather}
\sigma_{x'x'}  = -2\nu\hat{\sigma}\left[m-\tilde{z}'\right],
\label{eq:sigma-xx}\\
\sigma_{y'y'}  = -\hat{\sigma}\left[m-2\tilde{z}' + 
m\frac{\tilde{z}'^2+n^2}{m^2+n^2}\right],
\label{eq:sigma-yy}\\
\sigma_{z'z'}  = -\hat{\sigma}\left[m- 
m\frac{\tilde{z}'^2+n^2}{m^2+n^2}\right],
\label{eq:sigma-zz}\\
\sigma_{y'z'}  = 
\hat{\sigma}\left[n\frac{m^2-\tilde{z}'^2}{m^2+n^2}\right],
\label{eq:sigma-yz}\\
\sigma_{x'y'} = \sigma_{z'x'} = 0,
\label{eq:sigma-xy-zx}
\end{gather}
where $\hat{\sigma}$ is the maximum stress at the contact surface,  $(\tilde{x}', \tilde{y}', \tilde{z}')^T=b^{-1}~ (x', y', z')^T$ are the 
dimensionless coordinates of the stress field and $m$ and $n$ are given by 
\begin{gather}
m = \frac{\tilde{z}'}{\sqrt{2}|\tilde{z}'|}
\bigg\{(1-\tilde{y}'^2+\tilde{z}'^2) \nonumber\\
+\sqrt{(1-\tilde{y}'^2+\tilde{z}'^2)^2 + 
4\tilde{y}'^2\tilde{z}'^2}\bigg\}^{\frac{1}{2}} ,\\
n = \frac{\tilde{y}'}{\sqrt{2}|\tilde{y}'|}
\bigg\{(\tilde{y}'^2-1-\tilde{z}'^2) \nonumber\\+ 
\sqrt{(1-\tilde{y}'^2+\tilde{z}'^2)^2 + 
4\tilde{y}'^2\tilde{z}'^2}\bigg\}^{\frac{1}{2}} .
\end{gather}

In the case of compressive stress, the sign of $\hat{\sigma}$ coincides with the sign of 
$\tilde{z}'$. The parameters $\nu=-S_{12}/S_{11} = 0.464$ and $Y = 1/S_{11} = 
23.99~\si{\GPa}$ (see Table~\ref{tab:mat-prop}) are Poisson's ratio and Young's 
modulus, respectively, and
\begin{equation}
b=2|\hat{\sigma}|R(\Theta_{\text{\cuo}} + \Theta_{\text{stressor}})
\end{equation}
is the half width of the contact area between stressor and crystal, with the 
stressor's radius $R$ and $\Theta = (1-\nu^2)/Y$. 

The line load $F$ in Fig.~\ref{fig:skizze} is related to the stress 
$\hat{\sigma}$ by  
\begin{equation}
F=\frac{\hat{\sigma}\pi b}{2} = \pi 
\hat{\sigma}|\hat{\sigma}|R(\Theta_{\text{\cuo}} + \Theta_{\text{stressor}}).
\end{equation}

Due to the elastic anisotropy of {\cuo}, the stress tensor $\bs(\br)$ has to be 
transformed into the crystal coordinates $x\hat{=}[100]$, $y\hat{=}[010]$ and 
$z\hat{=}[001]$ first, before the strain tensor $\be(\br)$ is calculated from it.
The components of strain tensor and stress tensor are then connected by the 
compliance constants
\begin{gather}
\epsilon_{xx} = S_{11}\sigma_{xx} +S_{12} (\sigma_{yy} + 
\sigma_{zz}),
\label{eq:stress-strain}\\
\epsilon_{yz} = S_{44}\sigma_{yz}/2
\end{gather}
for the diagonal and  off-diagonal components, respectively (the other 
components are given by cyclic permutations).

\begin{figure}
\includegraphics[width=\columnwidth]{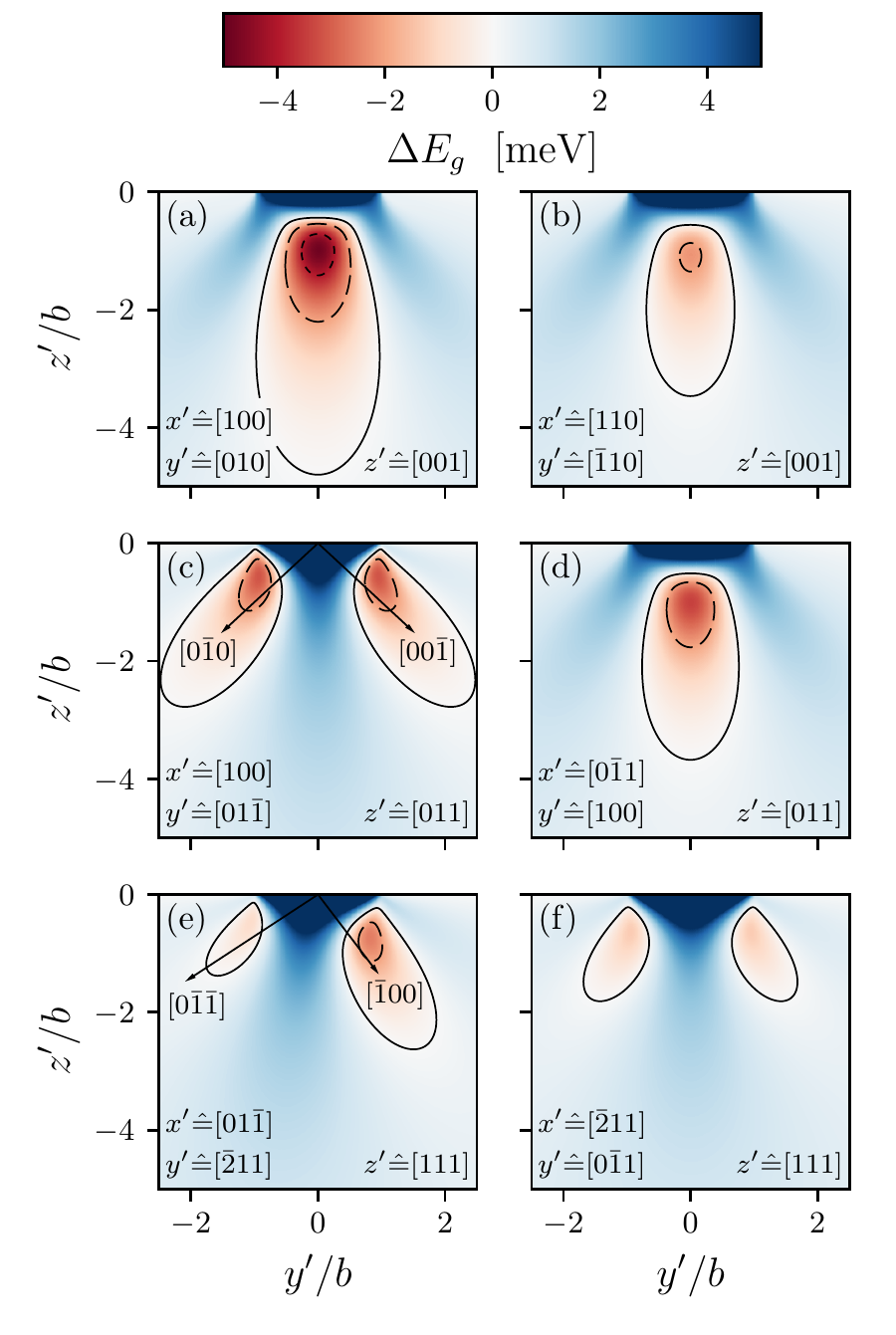}
\caption{Band-gap shifts $\Delta E_g$ for a maximum stress $\hat{\sigma} = 
-0.65~\si{\GPa}$ and different orientations of the stressor relative to the 
crystal axes. The contour lines mark the equipotential surfaces with $\Delta E_g 
= 0, -2, -4~\si{\milli\eV}$ (solid, dashed, dotted). This representation does 
not depend on the stressor's material properties, its radius or the line load, 
as long as $\hat{\sigma}$ is kept fixed.
\label{fig:eg}}
\end{figure}

Figure \ref{fig:eg} shows the band-gap shifts for different orientations of the 
crystal relative to the stressor and a maximum stress of $\hat{\sigma} = 
-0.65~\si{\GPa}$. The spatial splitting of the band-gap potential well for a 
main stress in $z'\hat{=}[011]$ direction viewed from $x'\hat{=}[100]$ 
(Fig.~\ref{fig:eg} (c)) has already been observed experimentally for spherical 
stressors \cite{naka2004}.
For the cases shown in Fig.~\ref{fig:eg} (a), (b) and (d), the potential is 
well approximated by a harmonic potential in $y'$-direction, and is 
Morse-like along the $z'$-direction.
\begin{figure}
\includegraphics[width=\columnwidth]{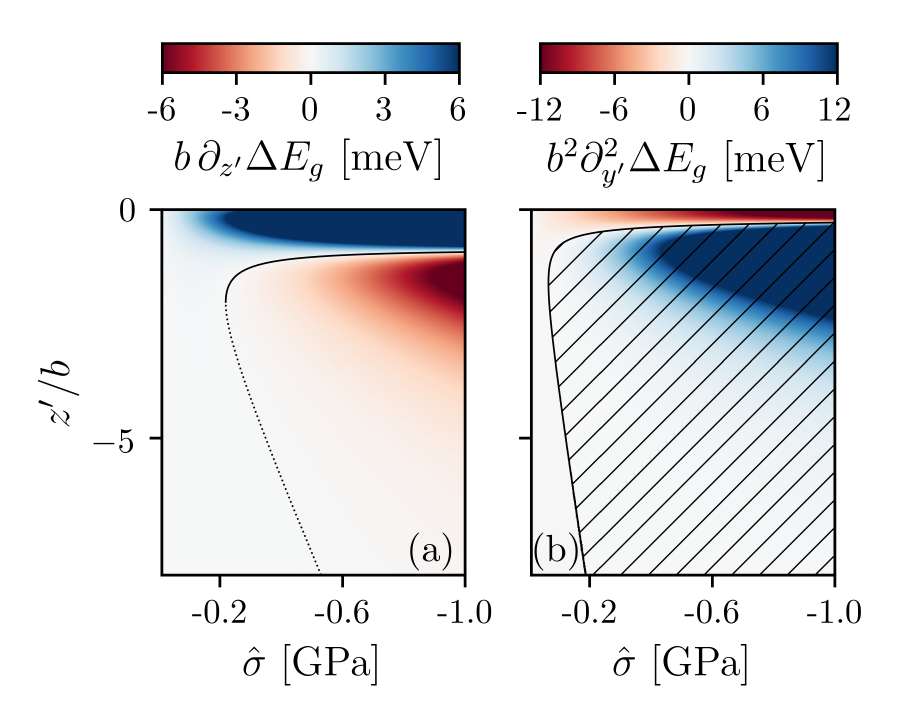}%
\caption{Derivatives of $\Delta E_g$ for the geometry of Fig. \ref{fig:eg} (a) 
and $y'=0$. (a): Potential gradient in $z'$ direction . The black lines mark the 
potential extrema, the upper branch (solid line) corresponds to the potential 
minimum.
(b): Second derivative in $y'$ direction. For symmetry reasons, the first 
derivative does always vanish in this configuration. $\Delta E_g$ has  a minimum 
in $y'$-direction in the blue (hatched) region and a maximum in the red 
region.\label{fig:deriv}}
 \end{figure}
 
Figure~\ref{fig:deriv} shows the range of stresses in which a potential minimum 
can form for the arrangement of Fig.~\ref{fig:eg} (a). The limiting factor for 
the formation of a trap is the minimum in the $z'$-direction, which only forms 
for stresses $|\hat{\sigma}|\gtrsim 0.25~\si{\GPa}$. 
%When interpreting the plots in Fig.~\ref{fig:deriv}, recall that $b\propto\hat{\sigma}$
%and  $\partial_{z'}= b^{-1}\partial_{\tilde{z}'}$ as well as $\partial^2_{y'} =b^{-2}\partial^2_{\tilde{y}'}$.
From the second derivatives at the trap minimum 
$\br_{min}$, the trap frequencies can be calculated as
\begin{equation}
\omega_{x_i} = \sqrt{\frac{1}{M}\left.\frac{\partial^2 
\Delta E_g(\br)}{\partial x_i^2}\right|_{\br=\br_{\min}}}
\end{equation}
where $M=(A_4^{-1} + (2B_1-A_1)^{-1})~m_e=1.55~m_e$ is the exciton mass and 
$x_i=y',z'$.
 
\begin{figure}
\includegraphics[width=\columnwidth]{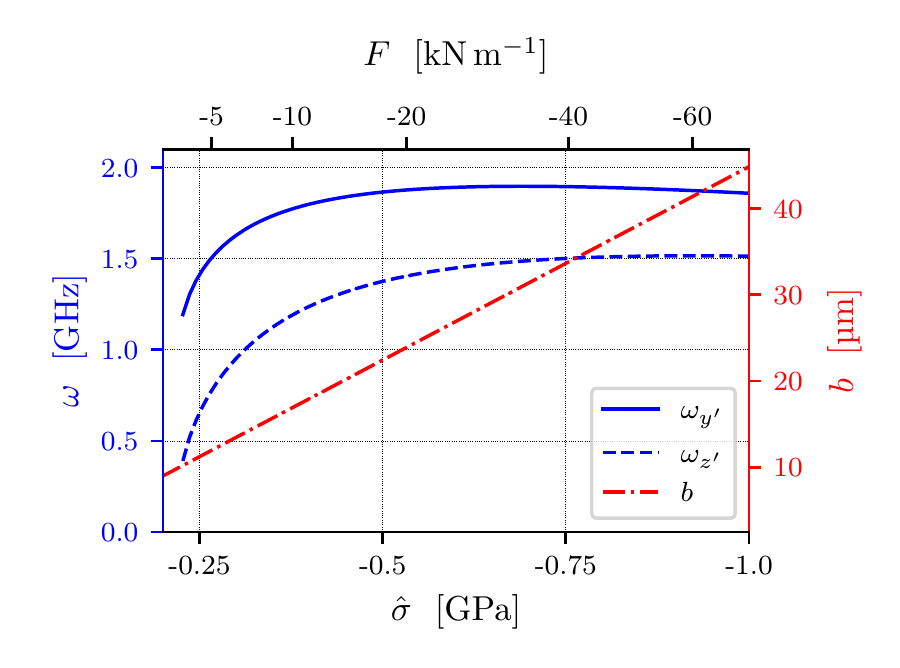}
\caption{Trap frequencies $\omega$ and contact half width $b$ for a glass 
stressor with $R=0.5~\si{\milli\meter}$, $Y=75~\si{\GPa}$ and $\nu = 0.3$ in 
the arrangement of Fig.~\ref{fig:eg} (a).\label{fig:frequencies}}
\end{figure}

The trap frequencies for a stressor with $R=0.5~\si{\milli\meter}$, 
$Y=75~\si{\GPa}$ and $\nu = 0.3$ are shown in Fig.~\ref{fig:frequencies}. For 
fixed $\hat{\sigma}$, the frequencies scale as 
$\omega\propto b^{-1}\propto R^{-1}
(\Theta_{\text{\cuo}}+\Theta_{\text{stressor}})^{-1}$.  The results in Fig. 
\ref{fig:deriv} have been obtained by a complete 
diagonalisation of $\oh_r(0, \be)$, as the perturbation theoretical treatment 
introduces errors of  $\sim 2~\si{\milli\eV}$ at the trap minimum for 
$\hat{\sigma}=-1~\si{\GPa}$.

In order for the assumption of slowly varying strain to hold, the trap dimension 
should be much larger than the spatial dimension of the exciton states, i.e. 
$\langle r_{n,\ell} \rangle\ll b$. On the other hand, the excitonic lifetimes 
$\tau_{n,\ell}$ should be larger than the inverse trap frequencies 
$\omega^{-1}$ in order to have a meaningful definition of ``trapped excitons''.
The inverse trap frequencies in our numerical example are on the same order of 
magnitude as the lifetimes of strain-free the Rydberg excitons 
$\tau_{n,\ell}$ that can be estimated from the FWHM linewidths 
\cite{kazimierczuk2014, schweiner2016} $\hbar\gamma_{n,\ell}\approx \hbar / \tau_{n,\ell}$.
This results in lifetimes on the order of $\SI{100}{\pico\second}$ for the 15P states,
compared to $\omega_x^{-1}\approx \SI{500}{ps}$ for the parameters used in Fig. \ref{fig:frequencies}.
As the trap frequency is inversely proportional to the trap dimension $b$, it can be tuned
by the use of stressors with different radii. 
Additionally, strain is known to influence excitonic 
lifetimes in {\cuo} \cite{denev2002}. To the best of our knowledge no detailed 
studies on the influence of strain on the lifetimes of
Rydberg excitons have been performed yet, which we will leave for future work.

\begin{figure}
\includegraphics[width=0.8\columnwidth]{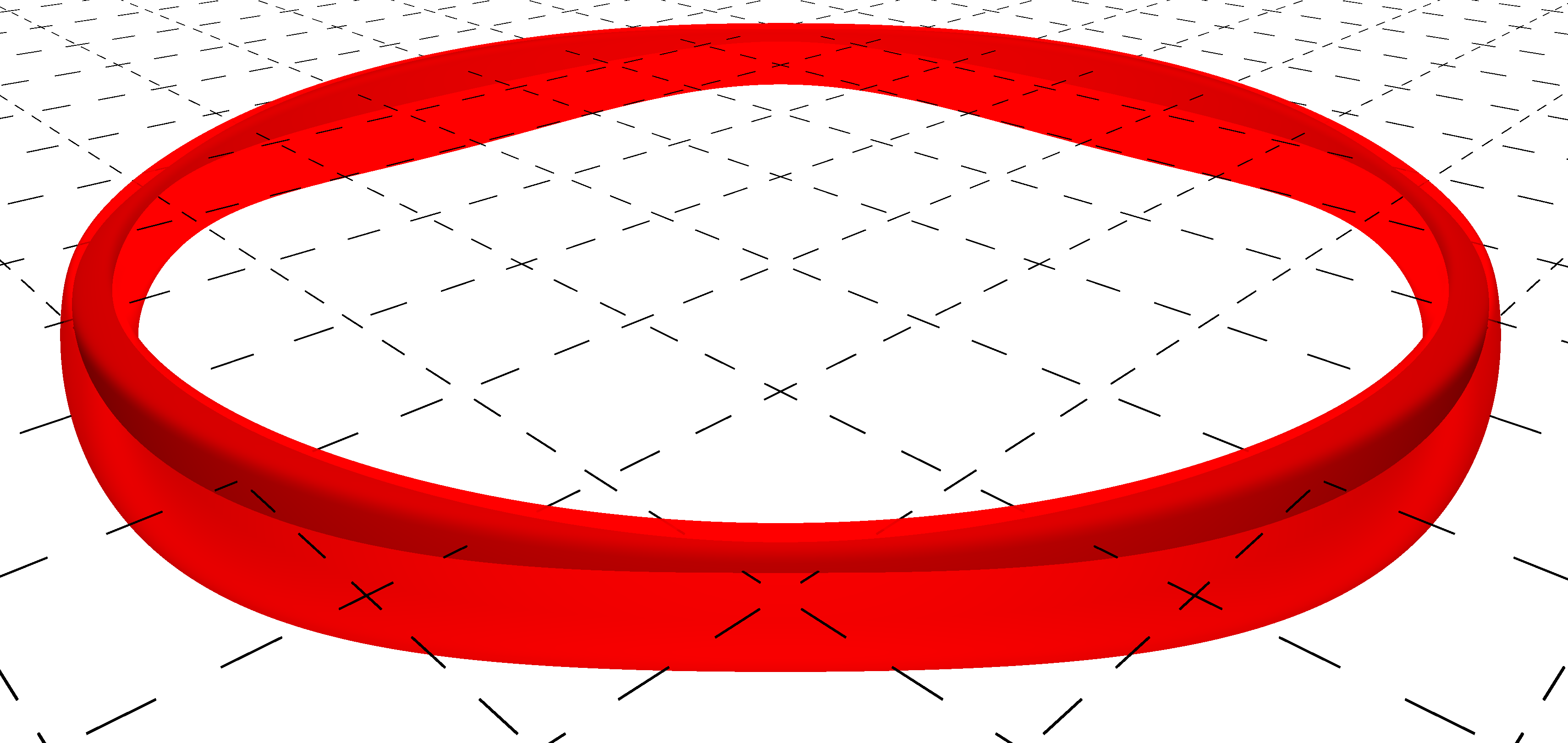}%
\caption{Isosurfaces for $E=0,-2~\si{\milli\eV}$ (outer and inner surface, 
respectively) of the strain potential induced by a toroidal stressor pressed 
onto a {\cuo} crystal along the $[001]$ axis with $\hat{\sigma} = 
-0.65\si{\GPa}$. The radius (center of torus to center of contact area) was set 
to $\rho = 20b$, the grid lines are separated by $5b$.\label{fig:torus}}
\end{figure}

In principle, much more intricate potential geometries can be created by 
corresponding stressors. As an example, Fig.~\ref{fig:torus} shows the 
approximate isosurfaces of the potential induced by a toroidal stressor pressed 
onto a {\cuo} crystal along the $[001]$ crystal axis and viewed along the 
$[110]$ direction. The isosurfaces were calculated under the assumption that 
the stressor is locally cylindrical. The modulation arises from the varying 
orientation of the stressor relative to the crystal axes (see 
Figs.~\ref{fig:eg} (a) and (b)).

\section{Influence of strain on the binding energies of Rydberg excitons}
\label{sec:binding-energies}

As the excitonic states of interest in this study are odd-parity Rydberg 
states, the exchange interaction (which does not affect odd-parity states) and 
the coupling to the green exciton series via the kinetic Hamiltonian 
(\ref{eq:hr}) (which should mostly affect states with large momentum space 
extension, i.e. small principal quantum number $n$) will be ignored.
This enables the calculation of the excitonic binding energies by solving the 
anisotropic Wannier equation of a spinless two-band model
\begin{equation}
\left[T(\bk,\be) + V_{e-h}(\bk)\star\right] \phi(\bk) = E \phi(\bk) 
\label{eq:aniso-wannier}
\end{equation}
with the convolution operator of the momentum-space Coulomb 
potential \cite{szmytkowski2012,schoene2016_prb} $V_{e-h}(\bk)$
\begin{equation}
V_{e-h}(\bk)\star\phi(\bk) = 
\frac{1}{(2\pi)^{3/2}} \int d^3\bk' V_{e-h}(\bk - \bk') \phi(\bk')
\end{equation}
and one eigenvalue $T(\bk,\be)$ of $\oh_r(\bk,\be)$ corresponding to a product 
state of a hole in a $\sy{7}$ state with an electron in a $\sy{6}$ state.
The approach to the solution of Eq. (\ref{eq:aniso-wannier}) used in this study 
is discussed in Appendix~\ref{sec:num-sol}.

\begin{figure}
\includegraphics[width=\columnwidth]{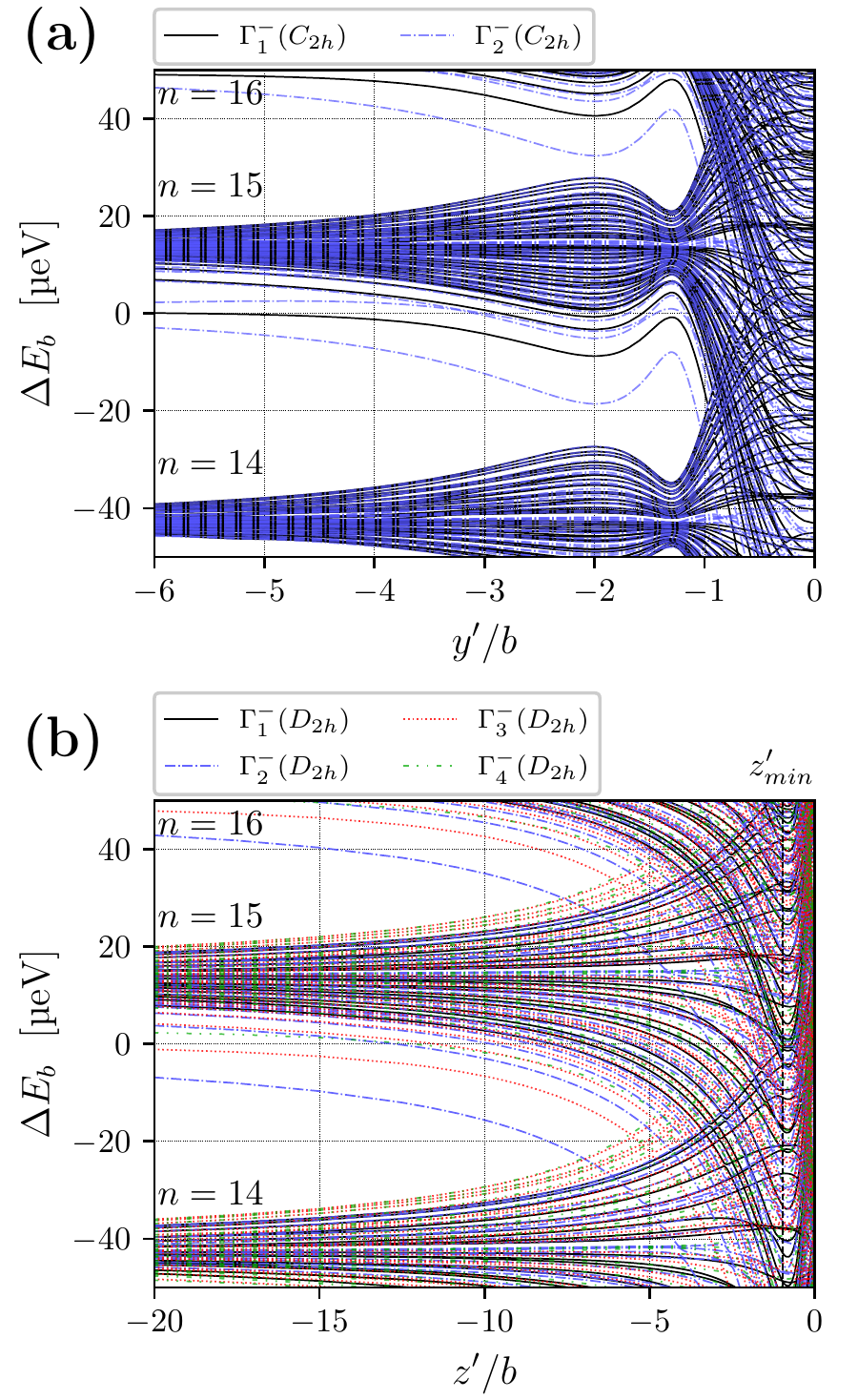}
\caption{Binding-energy shifts $\Delta E_b$ of the odd-parity excitons with 
$n=14,15$ and $16$ relative to the strain-free $15P$ resonance ($E_b = 
-397~\si{\micro\eV}$) for $\hat{\sigma}= -0.65~\si{\GPa}$ and the geometry of 
Fig. \ref{fig:eg} (a). The energies are sampled along lines through the 
potential minimum parallel to the $y'$- and $z'$-axis, 
respectively. (a): $z' = z'_{min} = -0.95\,b$, (b): $y'=0$.\\
The colors and linestyles denote the irreducible representations of the respective states and are the same as in Fig. \ref{fig:2P}.}\label{fig:15P}
\end{figure}

In what follows, we will assume that the coordinates of the stress field $x'$, 
$y'$ and $z'$ coincide with the crystal axes $x$, $y$ and $z$ (see 
Fig.~\ref{fig:eg} (a)).
At $y=0$, the strain becomes biaxial along $y$ and $z$ which reduces the local 
symmetry to $D_{2h}$ \cite{bir1974}. This does potentially lift all 
degeneracies in the spin-less two-band model as all irreducible representations 
of the single group are one-dimensional.
For $y\ne 0$, the strain is still confined to the $(y,z)$-plane (i.e. 
$\epsilon_{xx} = \epsilon_{xy} = \epsilon_{zx} = 0$), but the local symmetry 
reduces even further to $C_{2h}$. For other orientations of the stressor 
relative to the crystal axes, the local symmetries might be different. The only 
symmetry that is conserved under all (spatially homogeneous) strains is the 
inversion symmetry. Far from the contact zone $|y'| \gg b \vee 
|z'| \gg b$, the strain field vanishes and the $O_h$ symmetry is 
recovered.

\begin{figure}
\includegraphics[width=\columnwidth]{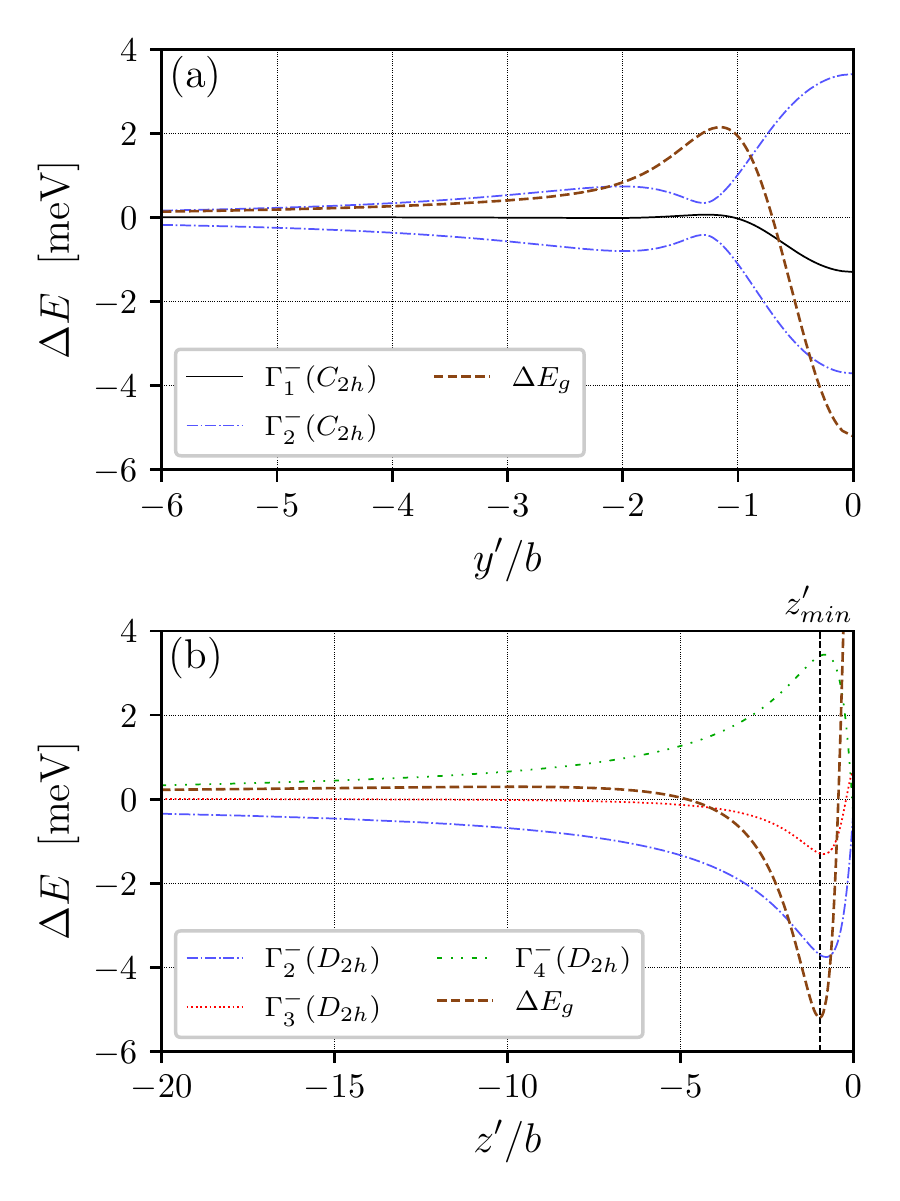}
\caption{Binding-energy shifts $\Delta E_b$ of the $2P$ resonances relative to 
the strain-free case ($E_b = -24.57~\si{\milli\eV}$) for the same conditions and 
with the same coding as in Fig. \ref{fig:15P}. The brown (dashed) line 
represents the band-gap shifts $\Delta E_g$ for comparison.\label{fig:2P}}
\end{figure}

Figure \ref{fig:15P} shows the binding energy shifts calculated from the 
two-band model including all odd-parity basis functions up to $\ell_{max}=25$ and $n_{max}=30+\ell$.
The strain-induced shift in the binding energy $\Delta E_b$ is on the order of 
$\si{\micro\eV}$ for the Rydberg states. As the band-gap shift is on the order 
of $\si{\milli\eV}$, it is the dominant potential contribution for states with 
large $n$: $\Delta E(\br)\approx \Delta E_g(\br)$. The energetical distance
between adjacent states might become important if one tries to address specific
states inside the trap. Due to finite laser spot-sizes, spectra will 
always contain a spatial average over the strain shifts in the region of the laser
spot.

For states with lower $n$, $\Delta E_b$ might reach the same order of magnitude 
as $\Delta E_g$. This is shown in Fig.~\ref{fig:2P}, which compares the 
shifts of the 2P resonances for the same conditions as in Fig.~\ref{fig:15P}. In the region
of the minimum of $\Delta E_g(\br_{min}) = \SI{-5.2}{\milli\eV}$ , $\Delta E_b$ reaches values
of $\SI{-3.7}{\milli\eV}$ for $\sy[-]{2}(D_{2h})$, $\SI{-1.3}{\milli\eV}$ 
for $\sy[-]{3}(D_{2h})$ and $\SI{3.4}{\milli\eV}$ 
for $\sy[-]{4}(D_{2h})$.
The positions of the minima of the total potential $\Delta E_b + \Delta E_g$ differ slightly for
the different symmetries, varying from  $z' = -0.94~b$ for $\sy[-]{2}(D_{2h})$ to $z' = -1.0~ b$ for
$\sy[-]{4}(D_{2h})$.

\section{Conclusion and Outlook}\label{sec:outlook}

We have shown that one can exploit the band-gap modification in a crystal to 
generate strain waveguides for Rydberg excitons that have similar or even 
superior parameters to optical dipole traps for atoms. For example, trap 
frequencies up to GHz and trap depths on the order of meV can potentially be 
reached. This implies that similar trapping geometries than those developed for 
atoms can be produced for excitons. The choice of stressor, together with the 
crystal orientation, determines the precise shape of the trapping potential 
with the possibility to create complex potential landscapes.

Our calculations have shown that, if one is only interested in the excitonic 
strain potential, it suffices to consider only the band-gap shift $\Delta E_g$ 
for the Rydberg states as it is about two orders of magnitude larger than the 
shift of the binding energies. The Rydberg states do, however, show a strong
coupling, and potentially chaotic motion, in regions of high strain. Moreover, 
strain is known to influence the lifetime of excitons which will be the subject 
of future work.

\begin{acknowledgments}
We acknowledge support by the Deutsche Forschungsgemeinschaft (DFG) via the 
Focussed Research Programme (SPP) 1929 'Giant Interactions in Rydberg systems'.
\end{acknowledgments}
\appendix

\section{Numerical solution of the anisotropic two-band model}
\label{sec:num-sol}

We will briefly review the numerical method used to compute the solutions of 
the isotropic, nonparabolic Wannier equation \cite{schoene2016_prb}, and 
extend it to anisotropic band structures. Starting point is the anisotropic 
momentum-space Wannier equation (\ref{eq:aniso-wannier}) written as a Sturmian 
eigenvalue problem
\begin{equation}
\left[\frac{\hbar^2\bk^2}{2\mu} +\Delta T (\bk)-E\right] \Psi(\bk) =- 
\lambda(E)~V_{e-h}(\bk)\star\Psi(\bk) \label{eq:aniso-wannier-sturmian}
\end{equation}
with the anisotropic, nonparabolic part of the hole dispersion $\Delta T(\bk)$ 
and the reduced effective electron-hole mass $\mu$, which both might depend on 
the strain tensor $\be$. In this calculation, however, the strain field is 
assumed to be constant over the extension of the exciton and those dependencies 
are not considered explicitly.

The Sturmian eigenvalues $\lambda(E)$ mark the excitonic eigenenergies 
$E_{n,\Gamma}$ by the condition $\lambda(E_{n,\Gamma}) = 1$. Hence, there is a 
parametric dependence of Eq.~(\ref{eq:aniso-wannier-sturmian}) on the sought 
energy eigenvalue $E$ which can be solved for the $\lambda(E)$ and the 
excitonic eigenenergies retrieved from them.

In contrast to Ref.~\cite{schoene2016_prb}, $\Delta T (\bk)$ is no longer
assumed to be isotropic, but may transform according to $\sy{1}$ of $O_h$ 
without strain and may retain only the inversion symmetry for arbitrary strain. 
In the strainless case, the exciton states split according to S$\mapsto 
\sy{1}$, P$\mapsto\sy[-]{4}$, D$\mapsto \sy{3}\oplus\sy{5}$ and 
F$\mapsto\sy[-]{2}\oplus\sy[-]{4}\oplus\sy[-]{5}$ in this 
model \cite{koster1963}.
Therefore, trial functions with the correct symmetries can be constructed as
\begin{equation}
\Psi_{\Gamma_j}(\bk)=\sum_{\ell_\alpha}~F^{\Gamma_j}_{\ell_\alpha}(k)~\psi^{
\Gamma_j}_{\ell_\alpha}(\bn_k),\label{eq:expansion}
\end{equation}
where $\psi^{\Gamma_j}_{\ell_\alpha}(\bn_k)$ is the $\alpha$th cubic 
harmonic \cite{vdLage1947} of order $\ell$ that transforms like a basis 
function $\Gamma_j$ of the irreducible representation $\Gamma$, and 
$F^{\Gamma_j}_{\ell_\alpha}(k)$ is the corresponding radial wave function.
In terms of the spherical harmonics $Y_\ell^m(\bn_k)$, the relevant real-valued 
cubic harmonics up to fifth order for $\sy[-]{4,z}$ and tenth order for $\sy{1}$ 
are
\begin{gather}
\psi^{\sy{1}}_{0_1} = Y_0^0,\label{eq:g1-0}\\
\psi^{\sy{1}}_{4_1} = \sqrt{\frac{5}{24}}(Y^4_4 + Y^{-4}_{4}) + 
\sqrt{\frac{7}{12}}Y^0_4,\\
\psi^{\sy{1}}_{6_1} = \sqrt{\frac{7}{16}}(Y^4_6 + Y^{-4}_{6}) - 
\sqrt{\frac{1}{8}}Y^0_6,\\
\psi^{\sy{1}}_{8_1} = \sqrt{\frac{65}{384}}(Y^8_8 + Y^{-8}_{8}) 
+\sqrt{\frac{7}{96}}(Y^4_8 + Y^{-4}_{8})\nonumber\\
+ \sqrt{\frac{33}{64}}Y^0_8,\\
\psi^{\sy{1}}_{10_1} = \sqrt{\frac{187}{768}}(Y^8_{10} + Y^{-8}_{10}) 
+\sqrt{\frac{11}{64}}(Y^4_{10} + Y^{-4}_{10})\label{eq:g1-10}\nonumber\\
 - \sqrt{\frac{65}{384}}Y^0_{10},\\
\psi^{\sy[-]{4,z}}_{1_1} = Y^0_{1},\\
\psi^{\sy[-]{4,z}}_{3_1} = Y^0_3,\\
\psi^{\sy[-]{4,z}}_{5_1} =  Y^0_5,\\
\psi^{\sy[-]{4,z}}_{5_2} = \sqrt{\frac{1}{2}}(Y^4_5 + Y^{-4}_{5}).
\end{gather}
 
Inserting the expansion into Eq. (\ref{eq:aniso-wannier-sturmian}) while 
simultaneously expanding 
\begin{equation}
D(\bk):=\frac{2 \mu \Delta T(\bk)}{\hbar^2} = \sum_{\ell_\alpha''}~ 
D_{\ell_\alpha''}(k) 
~\psi^{\sy{1}}_{\ell_\alpha''}(\bn_k),\label{eq:bs-expansion}
\end{equation}
yields 
\begin{gather}
\sum\limits_{\ell_\alpha'}\left[\left(k^2+q^2\right)\delta_{\ell_\alpha, 
\ell_\alpha'}+  \sum\limits_{\ell_\alpha''}~D_{\ell_\alpha''}(k) 
~C^{\Gamma_j}_{\ell_\alpha, 
\ell_\alpha',\ell_\alpha''}\right]F^{\Gamma_j}_{\ell_\alpha'}(k)\nonumber\\
=\frac{2\lambda(q)}{\pi a_B k}~ \int\limits_0^\infty 
dk'~k'~F^{\Gamma_j}_{\ell_\alpha}(k')~Q_{\ell}\left(\frac{k^2+k'^2}{2kk'}\right) 
\label{eq:fredholm}
\end{gather}
after multiplication by $\psi^{\Gamma_j\dagger}_{\ell_\alpha}(\bn_k)$ and 
integration over the unit sphere (for details, see 
Ref.~\cite{szmytkowski2012}). Here, $Q_\ell(x)$ denotes the Legendre functions 
of the second kind, $q = \sqrt{2\mu |E|}/\hbar$ is the renormalised energy, 
$a_B =4\pi\varepsilon_0\varepsilon_r\hbar^2 / (\mu e^2)$ the excitonic Bohr 
radius and 
\begin{equation}
C^{\Gamma_j}_{\ell_\alpha, \ell_\alpha',\ell_\alpha''} = \oint d^2\bn_k~ 
\psi^{\Gamma_j\dagger}_{\ell_\alpha} (\bn_k)~ \psi^{\sy{1}}_{\ell_\alpha''} 
(\bn_k)~ \psi^{\Gamma_j}_{\ell_\alpha'}(\bn_k)
\end{equation}
the coupling coefficients of the cubic harmonics, which can be evaluated 
in terms of Wigner 3-j symbols. Equation~(\ref{eq:fredholm}) can be rewritten as
\begin{gather}
J^{\Gamma_j}_{\ell_\alpha}(k)=
\frac{2\lambda(q)}{\pi a_B}\sum\limits_{\ell_\alpha'}\int\limits_0^\infty 
dk'~\left[\bm{T}^{-1}_{\Gamma_j}(k',q)\right]_{\ell_\alpha,\ell_\alpha'} 
J^{\Gamma_j}_{\ell_\alpha'}(k')\nonumber\\
\times\frac{Q_{\ell}\left(\frac{k^2+k'^2}{2kk'}\right)}{\sqrt{k^2+q^2}
\sqrt{k'^2+q^2}} \label{eq:fredholm-2}
\end{gather}
where
\begin{equation}
J^{\Gamma_j}_{\ell_\alpha}(k) = k \sqrt{k^2+q^2}\sum_{\ell_\alpha'} 
\left[\bm{T}_{\Gamma_j}(k,q)\right]_{\ell_\alpha,\ell_\alpha'} 
F^{\Gamma_j}_{\ell_\alpha'}(k)
\end{equation}
and
\begin{equation}
\left[\bm{T}_{\Gamma_j}(k,q)\right]_{\ell_\alpha,\ell_\alpha'}=
\delta_{\ell_\alpha, \ell_\alpha'}+  
\sum\limits_{\ell_\alpha''}~\frac{D_{\ell_\alpha''}(k)}{k^2+q^2} 
C^{\Gamma_j}_{\ell_\alpha, \ell_\alpha',\ell_\alpha''}.
\end{equation}

As in the case of an isotropic band structure, the $k$-dependent part of the 
integral kernel in  Eq.~(\ref{eq:fredholm-2}) can be expressed as 
\cite{szmytkowski2012}
\begin{gather}
M_\ell(q,k,k') = \frac{2}{\pi 
a_B}~\frac{Q_{\ell}\left(\frac{k^2+k'^2}{2kk'}\right)}{\sqrt{k^2+q^2}\sqrt{
k'^2+q^2}} \nonumber\\
= \sum\limits_{n_r=0}^\infty g_{n_r,\ell}(k,q)~g_{n_r,\ell}(k',q)
\end{gather}
where
\begin{gather}
g_{n_r,\ell}(k,q) = \ell! \sqrt{\frac{n_r!}{\pi~a_B~(n_r+2\ell+1)!}}
\frac{(4qk)^{\ell+1}}{(q^2+k^2)^{\ell+3/2}}\nonumber\\
\times C^{(\ell+1)}_{n_r}\left(\frac{q^2-k^2}{q^2+k^2}\right)
\end{gather}
with the Gegenbauer polynomials $C^{(\alpha)}_{n_r}(x)$. The functions
$g_{n_r,\ell}(k,q)$ are orthogonal with respect to the radial quantum number 
$n_r$,
\begin{equation}
\int\limits_0^\infty dk~ g_{n_r,\ell}(k, q) g_{n_r',\ell}(k,q) = 
\delta_{n_r,n_r'}\frac{1}{q~a_B~ (n_r+\ell+1)},
\end{equation}
and they form a natural basis in which to expand the 
$J^{\Gamma_j}_{\ell_\alpha}(k)$ in order to solve Eq.~(\ref{eq:fredholm-2}) 
numerically.

In the case of strain, the calculation is equivalent with the exception that
the correct lattice harmonics have to be chosen as the angular expansion functions 
(e. g. those for the point groups  $C_{2h}$ and $D_{2h}$ for the $y'$- and $z'$-directions 
in Sec.~\ref{sec:binding-energies} or  $C_i$ in the most general case). These can
be constructed by comparing the coupling coefficients given by Koster \cite{koster1963}
and the Clebsch-Gordan coefficients.

For $D_{2h}$, the odd-parity lattice harmonics are (with odd $\ell$ and the 
obvious limits on $\alpha$)
\begin{gather}
\psi^{\sy[-]{1}}_{\ell_\alpha} = i\sqrt{\frac{1}{2}} \left(Y_\ell^{2\alpha} - 
Y_\ell^{-2\alpha}\right),\\
\psi^{\sy[-]{2}}_{\ell_\alpha} = i\sqrt{\frac{1}{2}} \left(Y_\ell^{2\alpha+1} + 
Y_\ell^{-(2\alpha+1)}\right),\\
\psi^{\sy[-]{3}}_{\ell_0} = Y_\ell^0,\\
\psi^{\sy[-]{3}}_{\ell_\alpha} = \sqrt{\frac{1}{2}} \left(Y_\ell^{2\alpha} + 
Y_\ell^{-2\alpha}\right),\\
\psi^{\sy[-]{4}}_{\ell_\alpha} = \sqrt{\frac{1}{2}} \left(Y_\ell^{2\alpha+1} - 
Y_\ell^{-(2\alpha+1)}\right).
\end{gather}
For $C_{2h}$ they can be expressed as 
\begin{gather}
\psi^{\sy[-]{1}}_{\ell_{\alpha_{\pm}}} = 
\sqrt{\pm\frac{1}{2}} \left(Y_\ell^{2\alpha} \pm Y_\ell^{-2\alpha}\right),\\
\psi^{\sy[-]{1}}_{\ell_{0}} = Y_\ell^{0},\\
\psi^{\sy[-]{2}}_{\ell_{\alpha_{\pm}}} = \sqrt{\mp\frac{1}{2}} \left(Y_\ell^{2\alpha+1} 
\pm Y_\ell^{-(2\alpha+1)}\right) ,
\end{gather}
if the $z$-axis is chosen as the twofold rotation axis. For $C_i$ every 
odd-parity function belongs to $\sy[-]{1}$.

The $\bk$-space extension of the Rydberg states scales roughly as $n^{-1}$. 
Therefore, the most important terms for highly excited states in 
(\ref{eq:bs-expansion}) are those with $D_{l_\alpha}(k)\propto k^2$ for 
$k\rightarrow 0$, which
is only fulfilled by terms of zeroth and second order for the $\sy{7}$ band. 
Including only those quadratic terms, the problem at hand can be expressed as 
the anisotropic Kepler problem
\begin{equation}
\left[\frac{\hbar^2 k_x^2}{2m_x}+\frac{\hbar^2 k_y^2}{2m_y}+\frac{\hbar^2 
k_z^2}{2m_z}+ V_{e-h}(\bk)\star\right] \phi(\bk) = 
E\phi(\bk).\label{eq:aniso-kepler}
\end{equation}
For $O_h$-symmetry, there are no terms with $\ell = 2$ that transform like 
$\sy{1}$, and the effective mass at the zone center of the $\sy{7}$ valence 
band is isotropic. Hence, the coupling between different $\ell$ is small and 
the ansatz (\ref{eq:expansion}) can be truncated at low $\ell_\alpha$. 
Inside the traps, the effective mass is quite anisotropic and states of 
different $\ell$ couple strongly. Therefore, all odd-parity basis functions with
$n_r = n - \ell -1 \le 29$ and $\ell \le  25$ had to be included, in order for the calculation of the 
binding-energy shifts of the $n=15$ resonances to converge (giving bases of up 
to $\sim 5500$ basis functions). As the number of basis states scales with 
$n^2$, the algorithm becomes computationally very expensive and not feasible 
for $n\gtrsim15$.

\bibliography{sources}
\end{document}